# Zeeman driven superconductor insulator transition in strongly disordered MoC film. STM and transport studies in transverse magnetic field.

M. Žemlička,[1,3,4] M. Kopčík,[1,5] P. Szabó,[1] T. Samuely,[2] J. Kačmarčík,[1] P. Neilinger,[3] M. Grajcar,[3] and P. Samuely[1,2]

[1] *Centre of Low Temperature Physics, Institute of Experimental Physics, Slovak Academy of Sciences, 04001 Košice, Slovakia*
[2] *Centre of Low Temperature Physics, Faculty of Science, P. J. Šafárik University, 04001 Košice, Slovakia*
[3] *Department of Experimental Physics, Comenius University, 84248 Bratislava, Slovakia*
[4] *Institute of Science and Technology Austria, am Campus 1, 3400 Klosterneuburg, Austria*
[5] *Faculty of Electrical Engineering and Informatics, Technical University, 04001 Košice, Slovakia*

Superconductor insulator transition in transverse magnetic field is studied in the highly disordered MoC film with the product of the Fermi momentum and the mean free path $k_F*l$ close to unity. Surprisingly, the Zeeman paramagnetic effects dominate over orbital coupling on both sides of the transition. In superconducting state it is evidenced by a high upper critical magnetic field $B_{c2}$, by its square root dependence on temperature, as well as by the Zeeman splitting of the quasiparticle density of states (DOS) measured by scanning tunneling microscopy. At $B_{c2}$ a logarithmic anomaly in DOS is observed. This anomaly is further enhanced in increasing magnetic field, which is explained by the Zeeman splitting of the Altshuler-Aronov DOS driving the system into a more insulating or resistive state. Spin dependent Altshuler-Aronov correction is also needed to explain the transport behavior above $B_{c2}$.

Two-dimensional (2D) superconductors can be driven through quantum superconductor-insulator transition (SIT) by tuning various physical parameters, such as disorder, voltage gating or magnetic field [1,2]. In general, two principal mechanisms of SIT exist. In bosonic scenario, the phase coherence of the Cooper-pair condensate is disrupted, puddles with variant superconducting order parameter emerge and survive even on the insulating side [3,4]. In fermionic scenario [5] the amplitude of the superconducting order parameter is fully suppressed driving the superconducting transition temperature to zero. Our previous papers demonstrated that increased disorder in ultrathin MoC films leads to the fermionic scenario of the suppression of superconductivity [6,7].

Recently, the exclusivity of the insulating ground state in 2D superconductors after quantum transition has been called into question. A quantum superconductor (anomalous) metal transition was suggested as a viable alternative [8,9]. In this study, it is impossible to distinguish these two options, due to finite temperatures. Hence, for the sake of clarity, we will address the effects simply as SIT in the following.

In this study we use another SIT control parameter – magnetic field oriented perpendicularly to the ultrathin MoC film. External magnetic field interacts with electrons via two different mechanisms: Zeeman and orbital couplings. For perpendicular fields, usually the latter mechanism is dominant. In our study, counter-intuitively, beside orbital coupling also Zeeman coupling plays an important role in the superconducting as well as the insulating/normal state. It is evidenced by a high upper critical magnetic field $B_{c2}$, by its square root dependence on temperature, as well as by the Zeeman splitting of the superconducting quasiparticle DOS measured by scanning tunneling microscopy (STM). This surprising finding is related to the fact, that the spin pair breaking is comparable to the orbital one, since the product of Fermi wavenumber and mean free path, known as the Ioffe-Regel parameter $k_F l$, is close to unity. In magnetic fields above $B_{c2}$ a logarithmic anomaly in the density of states due to the Altshuler-Aronov effect of 2D disordered metal [10,11] is present. Surprisingly, the anomaly is enhanced in further increased magnetic field driving the system deeper into insulating/resistive state. This to our knowledge never observed effect is also explained as a consequence of the Zeeman effect, now on the Altshuler-Aronov DOS. Zeeman spin-dependent Altshuler-Aronov correction is also needed to describe the temperature dependence of sheet resistances at low temperatures in fields above $B_{c2}$ strongly supporting the Zeeman driven SIT in transverse magnetic field in MoC.

MoC films of 3 nm thickness were prepared by the reactive magnetron sputtering onto a Si substrate as described in [12,13]. The Ioffe-Regel parameter $k_F l$ was determined for all our thin MoC films [6,12] from the free-electron-model. Transport and STM measurements were performed in magnetic fields up to 16 and 8 T, respectively, all at temperatures down to 0.3 K. Details are in Suppl. Mat. [14].

Strongly disordered ultrathin MoC films reveal quantum corrections to Drude conductivity with a negative temperature derivative of the sheet resistance $dR_\square/dT$ from room temperatures down to superconducting transition [6].

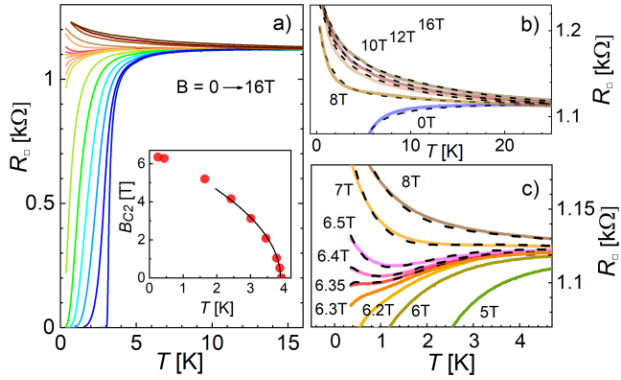

FIGURE 1. a) Temperature dependence of the sheet resistance of MoC film at magnetic fields $B$ = 0, 1, 2, 3, 4, 5, 6, 6.2, 6.3, 6.35, 6.4, 6.5, 7, 8, 10, 12 and 16 T. The inset shows temperature dependence of $B_{c2}(T)$ determined from resistive transitions (symbols). The solid black line is a $B_{C2}(T) \propto \sqrt{T_c - T}$ fit at $T \to T_c$. b) and c) show details of the resistive transitions, dashed black lines are fits (see the main text and Suppl. Mat. [14]).

Figure 1a) shows $R_\square(T)$ up to 16 T. At zero field a broad superconducting transition is observed due to superconducting fluctuations typical for disordered 2D systems. In magnetic field the initial drop of resistance is moving to lower temperatures and the transitions are further broadened. More details can be viewed in Fig. 1b) and c). In the field range 6.2 - 7 T, the slope of $R_\square(T)$ exhibits non-monotonous behavior. The onset of superconductivity is followed by a kink or minimum, the 6.2, 6.3 and 6.35 T curves exhibit a superconducting reentrant behavior with a second downturn of the $R_\square(T)$. Above 6.4 T no sign of reentrant superconductivity is observed indicating that the value is already above the upper critical field $B_{C2}$. At $B > 7$ T the slope of $R_\square(T)$ resumes the monotonous behavior in the whole temperature range with an insulator-like negative derivative $dR_\square/dT$ up to our highest field of 16 T. The inset of Fig. 1a) shows the temperature dependence of $B_{c2}(T)$ determined at 90 % of the "normal state" sheet resistance $R_\square^N(15\,K) = 1120\,\Omega$ (symbols) with a negative curvature that can be described by a $B_{C2}(T) \propto \sqrt{T_c - T}$ formula (solid line). Such curvature is typical for the critical field in paramagnetic limit $B^P$ [15], where pair breaking is caused by the Zeeman splitting. See also Suppl. Mat. [14].

The temperature dependence $R_\square(T)$ above $B_{C2} \sim 6.3$ T was fitted by the sum of the Drude conductivity, the contribution of superconducting fluctuations in the Cooper channel [11,16,17] and the Altshuler-Aronov correction in the diffusion channel [10,11]. As discussed in detail in Supplemental Material [14] a successful fit of the data is obtained only when the spin splitting effects are taken into account indicating that not only orbital but also the Zeeman effect is at play in the transverse magnetic field. Our tunneling experiments provide more conclusive arguments for a Zeeman driven SIT.

Figure 2a) displays typical tunneling spectra as a function of temperature. At the lowest temperature $T = 0.5$ K the spectrum shows reduced coherence peaks and in-gap states which thermal smearing alone cannot account for. Upon temperature increase the gap closes with the in-gap states growing and the coherence peaks ceasing. Above $T_c = 3.95$ K no superconducting features are visible. Still, the tunneling conductance shows a slight increase with the absolute value of voltage. The overall temperature dependence can be described within the thermally smeared Dynes DOS, $N_S^D(E) = (E + i\Gamma_D)/\sqrt{(E + i\Gamma_D)^2 - \Delta^2}$, with the superconducting gap $\Delta$ and the parameter $\Gamma_D$ responsible for spectral broadening [18,19]. The fitting curve at $T = 0.5$ K is plotted as a black solid line in Fig. 2a). The temperature dependence of the energy gap determined from the fit follows the BCS prediction with $\Delta(0) = 0.65$ meV, $\Gamma_D = 0.25$ meV, $T_c = 3.95$ K and $2\Delta/k_B T_c = 3.8$ in agreement with our previously published results [6,7]. The gapless superconducting DOS described by the Dynes modification of the BCS DOS [18] has been recently microscopically explained by the presence of local pair-breaking fields at arbitrary potential disorder [19]. In MoC such pair breaking fields are formed at the interface between the substrate and the film [7].

Figure 2b) shows an effect of perpendicular magnetic field on the tunneling spectra at $T = 0.5$ K. When $B$ is increased, the superconducting gap is gradually filled, and the coherence peaks are smeared. Notably, the gap structure of the tunneling spectra is not fully suppressed when $B_{c2}$ is approached but the tunneling conductance measured above $B_{c2} \sim 6.3$ T still shows a minimum at the zero-bias voltage. Astonishingly, this feature is enhanced by further increasing the field above $B_{c2}$. It is noteworthy that this effect is uniform across the sample surface, in contrast to the superconducting features as shown in the conductance maps of Fig. S1 in Suppl. Mat. [14]. It indicates that the spectra taken above 6.3 T do not reflect the superconducting DOS but rather the normal state properties without any spatial variations.

Figure 2c) displays a rare case of superconducting spectra with a smaller broadening parameter $\Gamma_D$. In magnetic fields of 2 T and 4 T this smearing enables the observation of two faint kinks marked by vertical bars. The effect is better resolved in the second derivative of the tunneling conductance $d^2I/dV^2$, where a pair of distinct minima is present. The distance between the minima is equal to $2\mu_B B$, consistent with Zeeman spin splitting.

Zeeman spin splitting and orbital pair-breaking effects are characterized by their respective critical magnetic fields. In a dirty type-II superconductor the orbital upper critical field $B_{c2} \sim \Phi_0/(\xi_0 l)$, where $\Phi_0$ is the flux quantum and $\xi_0 = \hbar v_F/\pi\Delta$ is the BCS coherence length. Due to the Zeeman coupling, the Cooper pairing would be destroyed by the Pauli depairing field $B^P \sim \Delta/\mu_B$, where $\mu_B$ is the Bohr magneton.

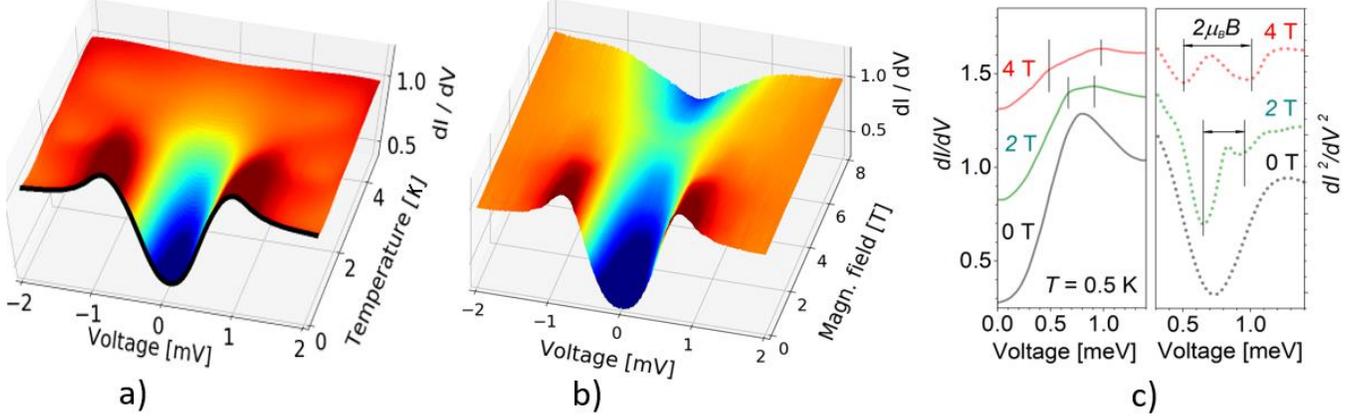

FIGURE 2. a) Temperature dependence of the STM differential conductance spectra in zero field. The black curve is a Dynes fit. b) Magnetic field dependence of the spectra at 0.5 K. c) Left: the spectra taken at $B = 0$ T (black), 2 T (green), and 4 T (red) magnetic fields with smeared Zeeman splitting features. Right: the second derivative of the spectra from the left panel. For description see the main text.

The ratio between these fields is $B^P/B_{c2} \sim k_F l$ showing that in materials which are close to SIT the Zeeman and orbital couplings are similar. The dominance of the Zeeman splitting of strongly interacting electrons was observed at the metal-insulator transition in disordered metals (e.g. the case of Si:B [20]). We assume that this effect can dominate also in superconductors near SIT. In the case of Zeeman splitting the spin-up and spin-down states in superconducting DOS are separated by $2\mu_B B \approx 0.116\, B$ [meV] in magnetic field $B$ and the gap peak splits in two [21]. It is documented in Fig. 2c) by the weakly smeared spectrum where the inequality $\Delta > \mu_B B > \Gamma_D$ holds.

It was shown in Ref. [19] that in Dynes superconductors the first order phase transition at the Zeeman driven critical magnetic field transforms into a continuous transition when $\Gamma_D$ exceeds $\sim 0.3\Delta(0)$. Still, there is a characteristic response to the magnetic field in tunneling DOS. While in case of orbital pair breaking the separation of the gap peaks shrinks, at Zeeman coupling the distance between the gap-like peaks remains nearly constant up to $B^P$. This means that gap filling rather than gap closing with increasing $B$ should be observed. Our MoC films are indeed Dynes superconductors [7,19] with spectral broadening in the range $\Gamma_D = 0.25\,\Delta - 0.36\,\Delta$. High spectral broadening in most of our spectra washes out the Zeeman splitting signatures, leaving just a broadened gap maximum. Still, one can observe in Fig. 3a) that the gap-peak position remains almost unchanged upon increasing $B$ up to $B_{c2}$ (dashed lines crossing the green curves) consistent with Zeeman coupling.

Thus, for the 3 nm MoC film in superconducting state, the effect of transverse magnetic field on the tunneling DOS, the square root temperature dependence of the upper critical field and its magnitude $B_{c2}(0) \sim 6.3$ T in the range of Clogston limit ($B^P$ [T] $\simeq 1.8 T_c$ [K], [22]) are evidences of the paramagnetic Cooper-pair breaking. Above $B_{c2}$ the Altshuler-Aronov anomaly in DOS is further enhanced by magnetic field.

In Fig. 3b) the normal state tunneling conductance taken at $B = 6.5$, 7 and 8 T fields is displayed in a wider voltage range. The inset plots the curves in the logarithmic scale. All the spectra reveal the same logarithmic energy dependence down to $\approx 1$ meV, where abrupt field dependent change occurs. In increased magnetic field the position of the change shifts slightly to higher energies and the zero-bias conductance (ZBC) decreases, similarly like in Fig. 2b).

It is well established, that a strong disorder enhances the electron-electron interaction. According to the Altshuler-Aronov (AA) theory of 2D systems [10,11] this is manifested by a logarithmic suppression of the DOS near the Fermi level at energies $k_B T \ll E \ll \Gamma$, where $\Gamma = \hbar/\tau \sim 10$ eV, inversely proportional to short mean free time $\tau$ in strongly disordered systems, is much greater than other relevant energy scales [23]. In our case the DOS starts to saturate at approx. 0.3 meV. This value is one order of magnitude higher than the thermal energy $k_B T = 0.043$ meV at $T = 0.5$ K, but is very close to the spectral broadening as discussed below.

Now we turn to the effect of magnetic field on the tunneling spectra above $B_{c2}$. Standard AA correction is field independent, but taking the electron spins into account in the diffusive particle-hole channel leads to a field dependence of the correction to the normalized DOS $N_N(E) = 1 - \delta\tilde{N}$. In the presence of finite relaxation rates, it can be derived from eq. 6.2a in Ref. [11] as

$$\widetilde{\delta N}(E,T,B) = \chi\Big\{f(\tilde{E},\tilde{\Gamma}_0) + \frac{\lambda_1}{2\lambda_0}\,[\,f(\tilde{E},\tilde{\Gamma}_1) + \frac{1}{2}\sum_{\alpha=\pm 1} f(\tilde{E} + \alpha \tilde{E}_Z, \tilde{\Gamma}_1)]\Big\}, \quad (1)$$

where arguments of the function $f$ are normalized to the thermal energy, as e.g. the Zeeman energy $\tilde{E}_Z = 2\mu_B B/k_B T$. The function $f$ is defined as

$$f(a,b) = -\frac{1}{2}\int_0^{\Gamma/k_B T} dx \frac{x}{x^2 + b^2} \frac{\sinh(x)}{\cosh(x) + \cosh(a)}.$$

The parameter $\chi$ includes constants as parameters of the tunneling barrier and the constant uncorrected density of states. The constants of $\lambda_0$ and $\lambda_1$ describe the interaction of an electron and a hole contribution of processes with total spin 0 and 1, respectively. The first and the second term in Eq. 1 originate from the interaction of an electron and a hole with a zero projection of the total spin on the direction of the magnetic field ($M = 0$), and the third and fourth terms, from interactions with $M = +1$ and $M = -1$, respectively. In the case when the ratio $\frac{\lambda_1}{2\lambda_0}$ is from the interval $\langle -1/3, 0 \rangle$ at zero magnetic field, the logarithmic anomaly of the first term is partially compensated by three equal logarithmic terms with opposite sign. In a finite field ($E_Z > 0$) a part of the electronic DOS at Fermi energy is removed due to spin polarization and two singularities/maxima should appear at $\pm E_Z$. Thus, due to the spin polarization in magnetic field the logarithmic singularity/minimum at the Fermi level is more pronounced than it was in the zero-field case. The parameters $\Gamma_0 = \Gamma_\varepsilon$ and $\Gamma_1 = \Gamma_\varepsilon + \Gamma_s$ responsible for the broadening of AA logarithmic singularity in DOS are determined by the energy relaxation rate $\Gamma_\varepsilon$ and the spin relaxation rate $\Gamma_s$ (see Eq. 2.31 in [11]).

We use the Eq. 1 to fit our field dependent tunneling conductance data from Fig. 2b) measured above $B_{c2}$. As the first step, the parameters characterizing the field dependent AA effect $\frac{\lambda_1}{2\lambda_0}$, $\chi$, $\Gamma_0$ and $\Gamma_1$ have been determined from the tunneling conductance measured deeply in the normal state at $B = 8$ T and $T = 0.5$ K. Afterwards for the fits at lower fields the same set of parameters was applied. Below $B_{c2}$, a formula which combines the Dynes DOS and the AA correction in the normal state has been used, in the form of $N(E) = N_S^D(E)N_N(\Omega)$. The latter term involves complex energy $\Omega(E) = \text{Re}[\sqrt{(E + i\Gamma_D)^2 - \Delta^2}$ necessary to keep the number of charge carriers constant after the transition to the superconducting state [24]. During the fit below $B_{c2}$ we varied the value of the energy gap $\Delta$ and slightly corrected the values of $\Gamma_D$. Since the spectra in Figs. 2b) and 3a) do not reveal any apparent spin splitting in the superconducting state due to large $\Gamma_D$, we use a simple Dynes formula, which at high spectral smearing is indistinguishable from the sum of two split Dynes DOS's.

As can be seen in Fig. 3a) our experimental tunneling spectra shown in cyan (above $B_{c2}$) and green curves (superconducting state) coincide well with the model (black lines). The values of the fitting parameters of the AA effect are $\frac{\lambda_1}{2\lambda_0} = -0.2$, $\tilde{\lambda}_0 = 0.145$. The broadening parameters $\Gamma_D$, $\Gamma_0$ and $\Gamma_1$ have about the same value $\approx 0.3$ meV, which is striking and deserves further investigations. The superconducting gap $\Delta(B)$ obtained from the fit follows a square field dependence (see Suppl. Mat. [14]), in agreement with the predictions of Herman and Hlubina [19] for the Zeeman splitting in Dynes superconductors with strong spectral smearing $\Gamma_D > 0.3\Delta$. Red points in Fig. 3c) represents the magnetic field dependence of the ZBC from Fig. 2b) and other spectra (not shown). The dash-dotted line depicts the contribution of the Dynes model, the dashed line is the spin dependent AA contribution and the solid line is the combined Dynes-AA model. It is obvious that the model provides a reasonable theoretical description of the measured tunneling spectra in both, the superconducting as well as the normal states.

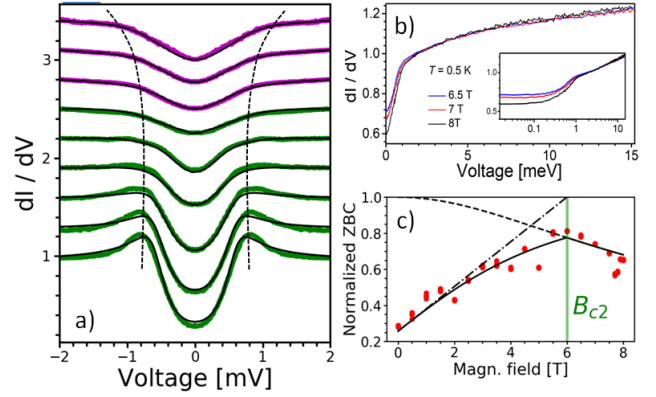

FIGURE 3. a) Tunneling spectra from Fig. 2b) in magnetic fields $B = 0, 0.5, 2, 3, 4, 6$ (superconducting state, green color) and at 6.5, 7 and 8 T (normal state, magenta) fitted to the model (black lines) described in the text. The dashed lines follow the field dependence of the superconducting gap maxima and the smeared Zeeman singularity positions. b) Tunneling spectra at $T = 0.5$ K and $B = 6.5$, 7 and 8 T (blue, red, black lines) measured in voltage window $(0, +15\text{mV})$. The inset of b) is the logarithmic plot of the same curves. c) Magnetic field dependence of the Zero Bias Conductance (red symbols) compared to theoretical models (dashed line - Altshuler-Aronov effect, dash-dot line - Dynes model, solid line – linear combination of the models).

The observation of the Zeeman driven SIT transition in homogeneously disordered thin films in perpendicular magnetic field is really surprising. This effect has not yet been observed possibly because there are only few experimental studies using local DOS measurements in such systems [25,26]. These works agree that the vortices are vanishing at strong disorder and the DOS in the normal state above $B_{c2}$ is strongly reduced at Fermi level. However, no systematic studies of the reduced normal state DOS have been done at fields above $B_{c2}$ in strongly disordered systems. In disordered NbN [26] the STM DOS measured in the vortex core (with presumably normal state DOS) features enhanced/deeper minimum in the reduced DOS than above

$T_c$ at $B = 0$. It is very similar to our case, but was neither noticed nor studied in detail. Notably, Wu et al. [27] have observed Zeeman splitting of the logarithmic AA DOS anomaly in the moderately disordered Al films in parallel field attributed not to the diffusive channel as in our MoC but to the Cooper channel. It is important to study the normal state properties of disordered superconductors at temperatures well below $T_c$ systematically.

In summary, the tunneling spectroscopy and transport measurements of highly disordered 3 nm thin MoC film with $k_Fl$ close to unity in transverse magnetic fields exhibit the Zeeman effects which play important role in both the superconducting state and the insulating state. In superconducting state the superconductivity is suppressed by the paramagnetic pair breaking as witnessed by the value of $B_{c2}$ close to the Clogston limit and also by the temperature dependence of $B_{c2}(T)$. The features of Zeeman coupling are found in the superconducting density of states where in increasing magnetic field up to $B_{c2}$ the superconducting gap is gradually filled rather than closed and in case of small broadening the Zeeman splitting of the gap-like peaks is observed. In normal state above $B_{c2}$ the experimental data can be even quantitatively described by the AA quantum corrections: the DOS shows enhancement of the logarithmic anomaly with increasing magnetic field driving the system deeper into insulating or resistive state. This effect, never observed experimentally before, is due to the spin dependent electron-hole interaction. Consistently spin dependent Altshuler-Aronov correction is also needed to explain the transport data measured above $B_{c2}$ in full agreement with the analysis of DOS. Thus, we can conclude that superconductor insulator transition in strongly disordered MoC film in transverse magnetic field is driven by the Zeeman effects. This transition not showing any emergent inhomogeneity in the spectral maps above $B_{c2}$ remains fully fermionic.

**Acknowledgement**

We gratefully acknowledge helpful conversations with B. L. Altshuler and R. Hlubina. The work was supported by the projects APVV-18-0358, VEGA 2/0058/20, VEGA 1/0743/19 the European Microkelvin Platform, the COST action CA16218 (Nanocohybri) and by U.S. Steel Košice.

[1] V. F. Gantmakher, V. T. Dolgopolov, Superconductor-insulator quantum phase transition, Phys. Usp. **53**, 1 (2010)

[2] N. Trivedi, in *Conductor Insulator Quantum Phase Transitions*, edited by V. Dobrosavljevic, N. Trivedi, and J.M. Valles, Jr. (Oxford University Press, Oxford, 2012), p. 329.

[3] M. P. A. Fisher, Quantum phase transitions in disordered two-dimensional superconductors, Phys. Rev. Lett.**65**, 923 (1990).

[4] B. Sacépé, C. Chapelier, T. I. Baturina, V. M. Vinokur, M. R. Baklanov, and M. Sanquer *et al*., Disorder-Induced Inhomogeneities of the Superconducting State Close to the Superconductor-Insulator Transition, Phys. Rev. Lett.**101**, 157006 (2008).

[5] A. M. Finkel'stein, Superconducting transition temperature in amorphous films, JETP Lett., **45**, 46 (1987).

[6] P. Szabó, T. Samuely, V. Hašková, J. Kačmarčík, M. Žemlička, M. Grajcar, J. G. Rodrigo, and P. Samuely, Fermionic scenario for the destruction of superconductivity in ultrathin MoC films evidenced by STM measurements, Phys. Rev. B **93**, 014505 (2016).

[7] V. Hašková, M. Kopčík, P. Szabó, T.Samuely, J. Kačmarčík, O. Onufriienko, M. Žemlička, P. Neilinger, M. Grajcar, P. Samuely, On the origin of in-gap states in homogeneously disordered ultrathin films.MoC case, Appl. Surface Sci.,**461**, 143 (2018).

[8] A. Kapitulnik, S. A. Kivelson, B. Spivak, *Colloquim*: Anomalous metals: Failed superconductors, Rev. Mod. Phys. **91**, 011002 (2019).

[9] Chao Yang et al., Intermediate bosonic metallic states in the superconductor-insulator transition, Science 10.1126/science.aax5798 (2019).

[10] B.L. Altshuler, A.G. Aronov, & P.A. Lee, Interaction Effects in Disordered Fermi Systems in Two Dimensions. Phys. Rev. Lett. **44**, 1288 (1980).

[11] B. L. Altshuler and A. G. Aronov, in *Electron-Electron Interactions in Disordered Systems*, edited by A.L. Efros, and M. Pollak, (Elsevier Science Publishers, North Holland, New York, 1985), Modern Problems in Condensed Matter Sciences **10**, 1 (1985).

[12] M. Trgala, M. Žemlička, P. Neilinger, M. Rehák, M. Leporis, Š. Gaži, J. Greguš, T. Plecenik, T. Roch, E. Dobročka, and M. Grajcar, Superconducting MoC thin films with enhanced sheet resistance, Appl. Surf. Sci. **312**, 216(2014).

[13] E.L. Haase, Preparation and Characterization of $MoC_x$ Thin Films, J. Low. Temp. Phys. **69**, 246 (1987).

[14] See Supplemental Material at http://link.aps.org/supplemental/

[15] Y. Matsuda and H. Shimahara, Fulde–Ferrell–Larkin–Ovchinnikov State in Heavy Fermion Superconductors, J.Phys. Soc. Jpn. **76**, 051005 (2007).

[16] A. Larkin, A. Varlamov, *Theory of Fluctuations in Superconductors*, Oxford University Press, New York (2005)

[17] V. Galitski and A. Larkin, Superconducting fluctuations at low temperature, Phys. Rev. B **63**, 174506 (2001).

[18] R. C. Dynes, V. Narayanamurti, and J. P. Garno, Direct Measurement of Quasiparticle-Lifetime Broadening in a Strongly-Coupled Superconductor, Phys. Rev. Lett. **41**, 1509 (1978).

[19] F. Herman, and R. Hlubina, Microscopic interpretation of the Dynes formula for the tunneling density of states, Phys. Rev. B **94**, 144508 (2016).

[20] S. Bogdanovich, P. Dai, M. P. Sarachik, and V. Dobrosavljevic, Universal Scaling of the Magnetoconductance of Metallic Si:B, Phys. Rev. Lett. **74**, 2543 (1995).

[21] R. Meservey, P. M. Tedrow and P. Fulde, Magnetic Field Splitting of the Quasiparticle States in Superconducting Aluminum Films, Phys. Rev. Lett. **25**, 1270 (1970).

[22] A. M. Clogston, Upper limit for the critical field in hard superconductors, Phys. Rev. Lett. **9**, 266 (1962).

[23] P. Neilinger, J. Greguš, D. Manca, B. Grančič, M. Kopčík, P. Szabó, P. Samuely, R. Hlubina, and M. Grajcar, Observation of quantum corrections to conductivity up to optical frequencies, Phys. Rev. B **100**, 241106 (2019).

[24] R. Hlubina, private communication

[25] I. Roy, R. Ganguly, H. Singh and P. Raychaudhuri, Eur. Phys. J. B 92, 49 (2019).

[26] Y. Noat, V. Cherkez, C. Brun, T. Cren, C. Carbillet, F. Debontridder, K. Ilin, M. Siegel, A. Semenov, H.-W. Hübers, and D. Roditchev, Phys. Rev. B 88, 014503 (2013).

[27] Wenhao Wu, J. Williams, and P. W. Adams, Zeeman Splitting of the Coulomb Anomaly: A Tunneling Study in Two Dimensions. Phys. Rev. Lett. **77**, 1139 (1996).

# Supplemental Material for „ Zeeman driven superconductor insulator transition in strongly disordered MoC film. STM and transport studies in transverse magnetic field"

M. Žemlička,[1,3,4] M. Kopčík,[1,5] P. Szabó,[1] T. Samuely,[2] J. Kačmarčík,[1] P. Neilinger,[3] M. Grajcar,[3] and P. Samuely[1,2]

[1]Centre of Low Temperature Physics, Institute of Experimental Physics, Slovak Academy of Sciences, 04001 Košice, Slovakia
[2]Centre of Low Temperature Physics, Faculty of Science, P. J. Šafárik University, 04001 Košice, Slovakia
[3]Department of Experimental Physics, Comenius University, 84248 Bratislava, Slovakia
[4]Institute of Science and Technology Austria, am Campus 1, 3400 Klosterneuburg, Austria
[5]Faculty of Electrical Engineering and Informatics, Technical University, 04001 Košice, Slovakia

## Correlation of superconducting and normal state properties with surface corrugations. Zero-bias conductance maps below and above $B_{c2}$

The STM experiments were carried out via a custom-made scanning tunneling microscope allowing experiments down to $T = 0.3$ K and magnetic fields up to $B = 8$ T [6,7]. At each point of the sample surface, first the topography value was acquired with the feedback loop on, then the feedback loop was turned off and the tunneling current was recorded while sweeping the bias voltage. High energy resolution is limited by thermal broadening given by ambient temperature as routinely verified by measuring the full energy gap of aluminum with $T_c$=1.2 K. Finally, by numerical differentiation we obtained a complete map of differential tunneling conductance spectra at each point. All spectra were normalized to their value at 2 mV.

A typical STM topography measured on 70 nm x 50 nm$^2$ surface of the 3nm thin MoC film is shown in Fig. S1a). The image was recorded with the set-point tunneling current $I_{set}$= 1 nA and the bias voltage $V_B$ = 50 meV at $T$ = 0.5 K. The surface with *rms* roughness of about 0.65 nm features characteristic approx. 6 nm wide boomerang-like structures. Atop the "boomerangs" we observe a distorted hexagonal lattice of atoms, typical for the (111) plane of a B1 lattice of the δ-phase of MoC [7].

Fig. S1b) shows the zero-bias tunneling conductance (ZBC) map at $T$ = 0.5 K and $B$ = 0 T extracted from the spectra acquired on the surface area presented in Fig. S1a). It illustrates a spatial variation of the superconducting properties and evidently correlates with the surface topography. This behavior is consistent for all conductance maps taken at bias voltages corresponding to energies inside the superconducting gap. For values outside the gap, we observed no correlation with the surface topography. Fig. S1c)-f) show the ZBC map of the same surface at the same temperature, but in magnetic fields 1, 5, 7 and 8 T, respectively. Also the ZBC maps at $B$ = 1 and 5 T (Fig. S1c) and d)) are sensitive to the surface corrugations. On the other hand, the maps measured at B = 7 and 8 T are spatially homogeneous with a virtually constant value of the ZBC, i.e. no correlation with the topography is found. Similar featureless maps were recorded at the field 6.5 T, but in a different surface area.

As shown in our previous paper [7] the superconducting properties of MoC films are sensitive to the surface morphology. In slightly thinner areas of the sample the tunneling spectra reveal higher spectral broadening $\Gamma_D$, smaller value of the superconducting gap Δ and higher in-gap conductance. This is due to the varying distance of the surface to the film-substrate interface with the pair-breaking fields. This is in steep contrast to our observation of the ZBC maps above 6.3 T. Hence, we can conclude that the logarithmic anomaly in DOS observed above 6.3 T is not connected with superconductivity and the sample is in the insulating/resistive state described by the quantum corrections due to Altshuler-Aronov (AA) effect.

We have to mention, that we did not observe superconducting vortices in our 3 nm samples. Brighter areas in images in Fig. S1 c) and d) taken at $B$ = 1 and 5 T with smaller energy gap could be places where the magnetic field penetrates the sample, but the vortex structure is not observed. The absence of vortices at fields below $B_{c2}$ was also reported in strongly disordered NbN thin films [25, 26]. Y. Noat *et al.* noticed [26], that the contrast needed to image vortices could be deteriorated also due to the fact that the tunneling spectrum inside the vortex core of strongly disordered samples had a dip comparable with the superconducting DOS. It is important to note, that the dip inside the vortex core (with presumably normal state DOS) reveals a similar reduction as the normal state DOS measured in our samples above $B_{c2}$ (see Figure 3b in the main manuscript)).

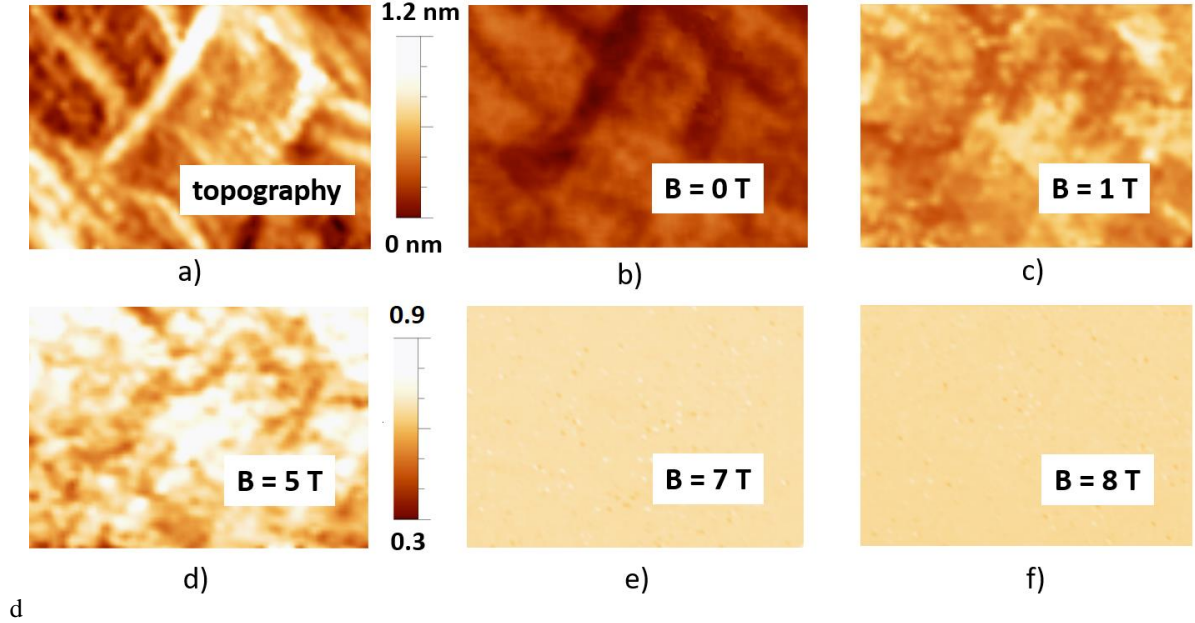

FIGURE S1. STM topography of a 70 nm x 50 nm surface a) and zero-bias conductance (ZBC) maps constructed from locally measured tunneling spectra at $T = 0.5$ K and different fields ($B = 0, 1, 5, 7$ and $8$ T in b), c) d) e) and f), respectively), in each point of the surface topography. The ZBC maps in b)-f) are presented in the same color scale as shown in d).

### Determination of $B_{c2}$ from the tunneling data:

The value of the upper critical magnetic field $B_{c2}(0)$ has been determined at $T = 0.5$ K from fitting the magnetic field dependence of the tunneling data, shown in Fig. 3a) of the manuscript, to the Dynes type DOS, taking into account the field dependent AA corrections, as well. The fitting procedure is described in the main text of the manuscript. The obtained magnetic field dependence of the energy gap shown in Fig. S2 follows the square root dependence $\Delta(B) \propto \sqrt{1 - B/B_{c2}}$ at fields $B \to B_{c2}$ in agreement with the theoretical predictions (see the main text of the manuscript) where $B_{c2}(0.5\text{ K}) \approx 6.3$ T.

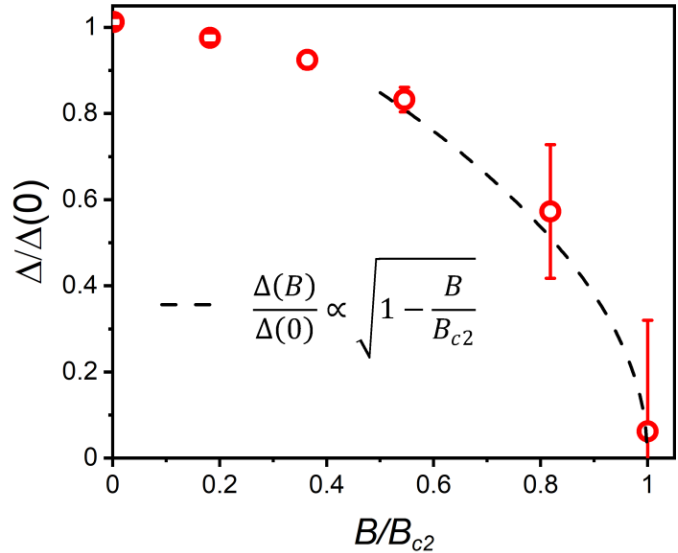

FIGURE S2. Magnetic field dependence of the superconducting energy dap determined from fitting of the tunneling data shown in Fig. 3a) of the manuscript (symbols). The energy gap values $\Delta$ and the applied magnetic fields $B$ are normalized to $\Delta(0) = 0.63$ meV and $B_{c2} = 6.3$ T. The dashed line represents the square root dependence.

# Fit of the transport data

## a) Upper critical magnetic field $B_{c2}$

As shown in the next section, the temperature dependence of the upper critical magnetic field $B_{c2}$ is an important parameter in fitting the temperature dependence of the sheet resistance. $B_{c2}$ has been determined from the $R_\square(T)$ resistive transitions shown in Fig. 1a) of the manuscript considering different criteria. Fig. S2 displays the resulting $B_{c2}(T)$ determined at 10%, 50%, 70% and 90% of the transition to the normal state, $R_\square^N(15\ \text{K}) = 1120\ \Omega$. The extrapolated $B_{c2}(0\ \text{K})$ is independent of the criterion and approaches a value of about 6.3 T. On the contrary, the zero field transition temperature $T_c(0\ \text{T})$ depends on the criterion and ranges from 3.2 to 3.9 K for 10% to 90 % of the transition, respectively. Importantly, the curvature of $B_{c2}(T)$ changes from positive to negative. As shown by the solid black line $B_{c2}(T)$ taken at 90 percent of the transition can be approximated by a negative curvature with $B_{C2}(T) \propto \sqrt{T_c - T}$. Such a curvature is typical for the critical field in paramagnetic limit $B^P$ [15], where pair breaking is due to Zeeman splitting. Similar $B_{C2}(T)$ with low temperature value of 6.3 T is obtained from our $R_\square(B)$ curves, measured at different fixed temperatures. These data will be published in our forthcoming paper.

As discussed below, the $B_{c2}(T)$ determined at 90% of the transition provides the best fit in the models which we use to explain the temperature dependence of the sheet resistance above $B_{c2}$. Moreover, such determined $B_{c2}(0.5\ \text{K})$ coincides well with the value obtained by the tunneling spectroscopy analyzed in the previous section.

It is important to notice, that the mechanism of pair-breaking can be characterized by the Maki parameter, defined as $\alpha = \sqrt{2} B_{c2}^{ORB}(0) / B_P(0)$, where $B_{c2}^{ORB}$ is the orbital and $B_P$ the paramagnetic critical magnetic field, respectively. It is known, that the Maki parameter is of the order of $\Delta(0)/E_F$ and its value is usually $\ll 1$ in classical BCS superconductors, where the orbital effects dominate. However, in materials with a heavy electron mass, or multiple small Fermi pockets $E_F$ can be small, $\alpha \geq 1$ and the paramagnetic pair breaking will dominate. High values of the Maki parameter have been observed in heavy fermion materials (e.g. *Y. Shinizu, et. al., J. Phys. Soc. Jpn. 80, 093701 (2011)*), in iron-based superconductors (for review see *G. Fuchs et. al., New J. Phys. 11, 075007 (2009)*), but not in strongly disordered systems like ours.

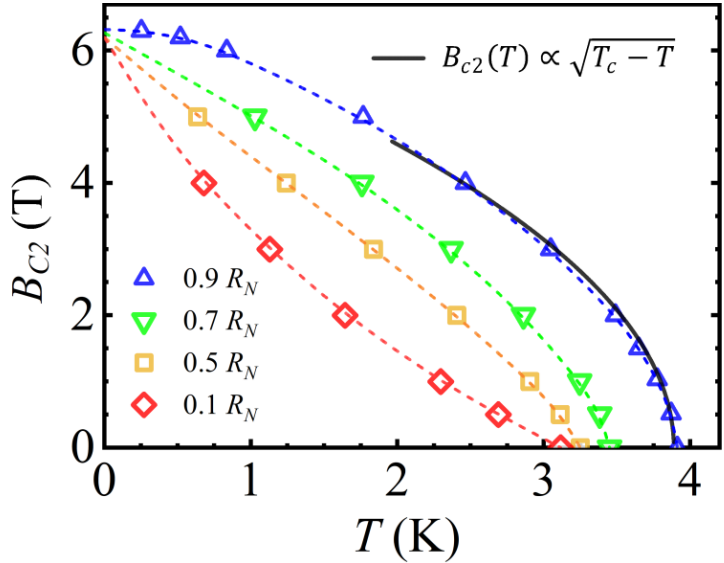

FIGURE S3. Temperature dependence of the perpendicular upper critical field $B_{c2}(T)$ determined at 10%, 50%, 70% and 90% of the normal state sheet resistance $R_\square^N(15\ \text{K}) = 1120\ \Omega$ (symbols). The dotted lines serve as a guide to the eye. The solid black line represents a $B_{C2}(T) \propto \sqrt{T_c - T}$ type of dependence at temperatures $T \to T_c$.

### b) Temperature dependence of the sheet conductance above $B_{c2}$

The total conductivity of a 2D disordered superconductor can be defined as the sum of the classical Drude conductivity $\sigma_0$ and the quantum corrections [1]

$$\sigma = \sigma_0 + \delta\sigma_{WL} + \delta\sigma_{SF} + \delta\sigma_{AA}, \tag{S1}$$

where $\delta\sigma_{WL}$ characterizes the weak localization correction, $\delta\sigma_{SF}$ and $\delta\sigma_{AA}$ are corrections connected with enhanced electron-electron interaction in the Cooper channel (superconducting fluctuations) and the diffusion channel (Altshuler-Aronov effect), respectively.

When perpendicular magnetic field is applied, the weak localization correction is quickly suppressed due to the time-reversal symmetry breaking [1]. Thus, $\delta\sigma_{WL}$ can be neglected at higher fields. In the low temperature limit $T \to 0$ and in magnetic fields above but close to $B_{C2}$ all corrections in the Cooper channel connected with the superconducting fluctuations (Aslamazov-Larkin, Maki-Thompson, density-of-states) are of the same order [1,16,17]. Their contribution can be described by a formula derived by Galitski and Larkin (GL) [17] as

$$\delta\sigma_{SF}(T \to 0) = \delta\sigma_{GL} = \frac{2e^2}{3\pi^2\hbar}\left\{-\ln\frac{r}{h} - \frac{3}{2r} + \psi(r) + 4[r\psi'(r) - 1]\right\}, \quad r = \frac{h}{2\gamma t}, \quad h = \frac{B - B_{C2}(T)}{B_{C2}(0)}, \quad t = \frac{T}{T_C(0)}, \tag{S2}$$

where $\gamma$ is Euler's constant and $\psi$ is the digamma function.

The Altshuler-Aronov (AA) correction in the diffusion channel of a 2D systems is expected to be field independent and can be expressed as [1, 11]

$$\delta\sigma_{AA} = -\alpha\frac{e^2}{\hbar}\ln\frac{T_C}{T^*}, \tag{S3}$$

where $T^*$ is an effective temperature and $\alpha$ is the weight of the AA contribution. We applied this model to describe our low temperature $R_\square(T)$ data from Fig. 1 (main text) measured at fields above $B_{C2}$, here represented as conductance $\sigma(T,H) = 1/R_\square(H,T)$. For the fitting, we used $\sigma_0 = 1/R_\square^N$ (Drude conductivity, where $R_\square^N(15\text{ K}) = 1120\ \Omega$) and the temperature dependence of $B_{C2}(T)$ (substituted to the GL term $\delta\sigma_{GL}$) determined at 10%, 50%, 70% and 90% of the resistive transition shown in Fig. S2 as input parameters. The best agreement was achieved with $B_{C2}(T)$ values determined at 90% of $R_\square^N$ and is displayed in Fig. S4a). In this manner, we were able to qualitatively describe the temperature dependence of the conductance $\sigma(T)$ only for magnetic fields $B$ = 6.35 T, 6.4 T and 6.5 T close to $B_{C2}$.

Altshuler and Aronov extended the quantum corrections to conductance by including spin interactions in magnetic field. The spin effects are usually insignificant in a homogeneous system in perpendicular magnetic field but our tunneling experiments suggest that in our sample they are relevant. In this case, the correction to Cooper channel in a 2D system is defined as (eq. 6.44 in [11])

$$\delta\sigma_{SF}(T,H) = \frac{e^2}{2\pi^2\hbar}\left\{\ln\frac{\ln\frac{k_BT_C}{\Gamma}}{\ln\frac{T_C}{T^*}} + \ln^{-1}\frac{T_C}{T^*}F_2\left[\frac{\tilde{E}_H}{2\pi}, \frac{\tilde{E}_Z}{\pi}, \frac{\tilde{\Gamma}_S}{\pi}\right]\right\}, \tag{S4}$$

where

$$\tilde{T}^* = \max\left\{1, \frac{\tilde{E}_H}{2\pi}, \frac{\tilde{E}_Z}{\pi}, \frac{\tilde{\Gamma}_S}{\pi}\right\}, \quad F_2(x_1, x_2, x_3) = x_1 \int_0^\infty \frac{\cos x_2 t}{\sinh x_1 t}\frac{e^{-x_3 t}}{\sinh^2 t}dt, \quad E_H = 3.6\,k_B T_C\frac{B}{B_{C2}}.$$

The energies $\tilde{E}_H, \tilde{E}_Z$ and the spin relaxation rate $\tilde{\Gamma}_S$ are normalized to thermal energy $k_B T$ (as in eq. (1) in the main text), $\Gamma = \frac{\hbar}{\tau} \sim 10$ eV is the unperturbed relaxation time [23].

The spin interaction effects will also modify the AA contribution to conductivity in the diffusion channel of a 2D system taking the form (eq. 6.47 in [11])

$$\widetilde{\delta\sigma_{AA}} = \widetilde{\delta\sigma_{AA}}(T) + \widetilde{\delta\sigma_{AA}}(B,T) =$$
$$= \frac{e^2}{4\pi^2\hbar}\left(\lambda_0^\sigma + \frac{3}{2}\lambda_1^\sigma\right)\ln\frac{1}{\tilde{\Gamma}} + \frac{e^2\lambda_1^\sigma}{4\pi^2\hbar}\int_0^\infty \frac{\partial^2(E\coth(\tilde{E}))}{\partial E^2}\left[\ln(\tilde{E} + \tilde{E}_Z) + \ln|\tilde{E} - \tilde{E}_Z| - 2\ln\tilde{E}\right]dE, \tag{S5}$$

where the first term in the sum $\widetilde{\delta\sigma_{AA}}(T)$ represents the temperature dependent part of the AA effect and the second one is the field dependent AA correction. Wavelets characterize again a normalization to $k_B T$ similarly as in the case of Zeeman energy

$\tilde{E}_Z = 2\mu_B B/k_B T$. Parameters $\lambda_0^\sigma$ and $\lambda_1^\sigma$ are interaction constants describing the interaction of an electron and a hole with total spin 0 and 1, respectively [11].

Now, we use the field dependent correction of the Cooper channel (S4) instead of the GL formula (S2) and the diffusion channel correction (S5) instead of AA term (S3) in the equation (S1). We fit again the temperature dependencies of the transport data above $B_{c2}$. As shown in Fig. S4c) this model describes our transport data (color lines) at zero field above $T_c$ and in high field range between 8T and 16 T (where the conductance decreases upon increasing field) not only qualitatively, but also quantitatively (see the black dashed lines). The fit parameters $\sigma_0 = 1/1058\ \Omega^{-1}$, $\lambda_0^\sigma = 2.681$ and $\lambda_1^\sigma = -1$ are field and temperature independent. However, the model fails for fields in the vicinity of $B_{c2}$. In this magnetic field range, the temperature dependence of the conductance displays a non-monotonous behavior showing first an onset of the superconducting transition which is not completed but at lower temperatures followed by the conductance decrease. We try to account for these effects by replacing the Cooper channel contribution (S4) by the Galitski-Larkin corrections (S2) and leave the AA correction (S5) with the previously obtained fit parameters $\sigma_0$, $\lambda_0^\sigma$ and $\lambda_1^\sigma$ unchanged. As shown in Fig. S4b) such a combination of the conductance corrections fits the data well. Again, the best agreement was achieved by using $B_{C2}(T)$ determined at 90% of the resistive transition as the parameters in (S2). Most importantly, a decent fit of the transport data above the upper critical field was achieved only when relation (S5) was employed, as demonstrated in Fig. S4b) and 3c). Therefore, we can conclude that the Zeeman spin interaction effects play the important role in the Altshuler-Aronov corrections to conductivity in the diffusion channel.

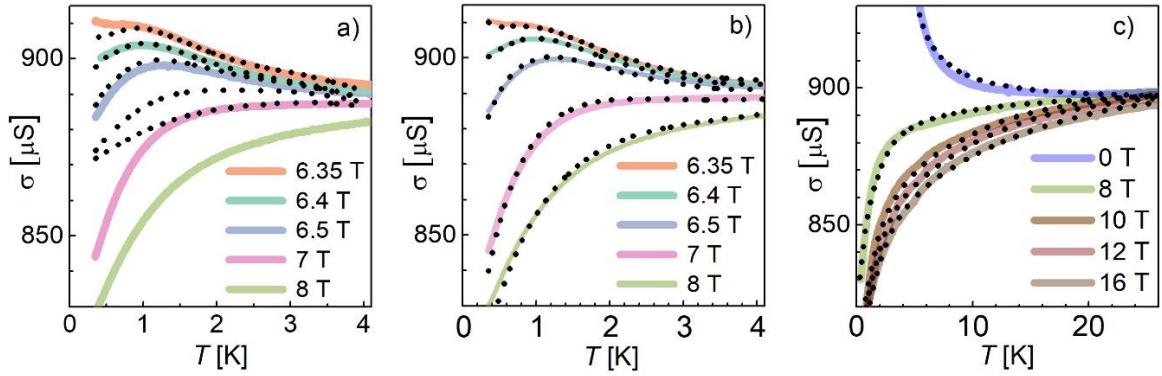

FIGURE S4. Conductance $\sigma$ vs. temperature $(T)$ above $B_{C2}(0)$ (solid colored lines). Black dotted lines represent the fitting curves to eq. (S1) with a) Galistki-Larkin (Cooper channel (S2)) and Altshuler-Aronov (diffusion channel (S3)) corrections, b) Galistki-Larkin (Cooper channel (S2)) correction and Altshuler-Aronov effect with spin interactions (diffusion channel (S5)) and c) Altshuler-Aronov corrections with spin interaction for Cooper channel (S4) and diffusion channel (S5).

It is also important to notice, that the ratio of the parameters used in the fit $\frac{\lambda_1^\sigma}{2\lambda_0^\sigma} = -0.19$ is almost the same, as in the fit of the normal state tunneling spectra to the spin dependent AA DOS correction, where $\frac{\lambda_1}{2\lambda_0} = -0.2$ (see the main text). It is another argument justifying the consistency of our approach. Finally, the analysis of the transport data measured in the fields up to 16 T strongly corroborates our findings obtained by scanning tunneling spectroscopy presented in the main text, showing that Zeeman effects are important for the suppression of superconductivity and enhancement of the insulating/resistive state in the transverse magnetic field applied to strongly disordered ultrathin MoC superconductor.